\documentclass[reprint, amsmath, amssymb, aps, pra, superscriptaddress]{revtex4-2}

\usepackage{graphicx} % Include figure files
\usepackage{dcolumn}  % Align table columns on decimal point
\usepackage{bm}       % bold math
\usepackage{hyperref} % add hypertext capabilities
\usepackage[mathlines]{lineno} % Enable numbering of text and display math
\usepackage{placeins} % For \FloatBarrier

\usepackage{tikz}
\usetikzlibrary{arrows.meta,positioning,calc}

\newlength{\tikzbbpad}
\tikzset{
  bbpad/.style={
    execute at end picture={
      \path (current bounding box.south west) ++(-\tikzbbpad,-\tikzbbpad) coordinate (bbSW);
      \path (current bounding box.north east) ++(\tikzbbpad,\tikzbbpad) coordinate (bbNE);
      \useasboundingbox (bbSW) rectangle (bbNE);
    }
  }
}

\begin{document}

\preprint{APS/123-QED}

\title{Payoff-Free Coordination Reveals Hidden Quantum Correlation Structure}

\author{Sinan Bugu}
	\email{sbugu@ncsu.edu}
	\email{sinanbugu@gmail.com}
\affiliation{Department of Physics and Astronomy, North Carolina State University, Raleigh, North Carolina, USA}
\affiliation{BuQuLab Research Laboratory, Winston-Salem, North Carolina, USA}

\date{\today}

\begin{abstract}
Coordination without communication is a recurring challenge in decentralized multi-agent systems. Classical shared-randomness strategies can generate correlated behavior, but we show that every anonymous classical common-cause strategy obeys a simple lower bound on pairwise coincidence: it cannot fall below the inverse alphabet size, regardless of how the shared randomness is constructed. In this work, we introduce a coordination protocol based on multipartite entanglement and demonstrate that it can violate this classical constraint. The resulting joint-action distributions exhibit nontrivial dependence while suppressing collisions, a regime that is inaccessible to classical symmetric strategies.  This separation appears without any advantage in expected payoff, indicating that entanglement reshapes the structure of coordination beyond what is captured by payoff-based metrics. The violation holds for $N < 2M$; with the four-action alphabet used here it survives up to seven agents and closes at eight.
\end{abstract}


\maketitle

% ==============================================================================
% SECTION I: INTRODUCTION
% ==============================================================================
\section{Introduction}
\label{sec:introduction}

Decentralized multi-agent systems often exhibit coordination patterns that cannot be inferred from any single agent's local rule. In many practical settings, agents cannot communicate, the coordination channel is constrained, and the environment imposes latent, round-by-round randomness that reshapes which joint actions are favorable or penalized. In such settings, the central scientific question is not whether a mechanism ``wins,'' but what \emph{correlation structure} it induces in the empirical joint-action distribution under identical external constraints.

In the restricted classical family studied here, dependence and collision are not separate knobs. If agents correlate by copying a shared latent value with probability $q$, the pairwise coincidence is $q^2(1-1/M) + 1/M$, which is monotone in $q$: every increment of dependence is bought with an increment of collision. In the limit of maximal correlation within this shared-latent family, agents achieve perfect dependence by outputting identical actions. While this maximizes information-theoretic metrics like total correlation, it is catastrophic for tasks requiring distributed coverage, load balancing, or anti-coordination, where local action collisions are penalized. Throughout, claims about an ``alignment trade-off'' refer to this explicit classical baseline family (defined in Sec.~\ref{sec:model}) rather than to arbitrary classical shared-randomness strategies.

Quantum platforms provide a distinctive route to break this trade-off. Multipartite entanglement can generate non-product joint statistics that sustain global dependence without forcing local alignment. In this work, we implement this via a \emph{Spontaneous Leader Election} protocol. Unlike idealized models that assume global knowledge of all measurement outcomes to resolve ties, our agents act solely on local measurement results. Each agent sees its own measurement bit and nothing else; no classical round of post-measurement reconciliation is permitted, even when the outcome is a tie or a vacancy. Such possibilities are discussed in the broader context of nonclassical correlations and nonlocality \cite{Bell1964, CHSH1969, BrunnerRMP2014, fritz2012beyond, navascues2020connector, Rosset2016, Acin2016, Wolfe2019Inflation, pearl2009causality, tavakoli2022bell}.

Recent literature on quantum coordination and quantum games has extensively explored nonlocality and entanglement as resources for distributed tasks \cite{Eisert1999, kolokoltsov2022dynamic, Natur2025, Bugu2020, bugu2025entanglement, ay2008information, Cuff2010}. However, these studies largely characterize quantum advantage in terms of payoff improvement or explicit Bell-inequality violations. The protocol we employ is deliberately not new: $W$ states are the canonical resource for anonymous leader election \cite{DHondt2006, Tani2012}, where they achieve symmetry breaking that is classically impossible with certainty. Our contribution lies elsewhere: we characterize the \emph{geometry of achievable joint-action distributions} that this known primitive induces under a latent environment and local noise, derive a coincidence lower bound obeyed by every anonymous classical common-cause strategy, and show that the quantum strategy violates it within a quantified window of team sizes, all in a regime where expected payoff provides no separation.

\begin{figure*}[tb]
\centering
\begin{tikzpicture}[
  bbpad,
  font=\sffamily,
  >=Latex,
  box/.style={draw=black, rounded corners=2pt, line width=0.6pt, inner sep=4pt, align=center},
  sbox/.style={draw=black, rounded corners=2pt, line width=0.6pt, inner sep=3pt, align=center},
  note/.style={font=\scriptsize, text=black!75, align=center},
  title/.style={font=\bfseries},
  arr/.style={->, line width=0.6pt},
  darr/.style={->, dashed, line width=0.6pt}
]

% Column positions (3 columns)
\coordinate (C1) at (0,0);
\coordinate (C2) at (6.2,0);
\coordinate (C3) at (12.4,0);

% Column titles
\node[title] at ($(C1)+(0,1.10)$) {Independent};
\node[title] at ($(C2)+(0,1.10)$) {Shared-latent};
\node[title] at ($(C3)+(0,1.10)$) {Quantum-correlated};

% Hidden field (one per column)
\node[box] (H1) at ($(C1)+(0,0.35)$) {Hidden field $F$};
\node[box] (H2) at ($(C2)+(-0.9,0.35)$) {Hidden field $F$};
\node[box] (H3) at ($(C3)+(0,0.35)$) {Hidden field $F$};

\node[note] at ($(C2)+(0,-0.28)$) {latent, unobserved};

% Shared resources
\node[note] (none) at ($(C1)+(0,-0.95)$) {no shared\\resource};

\node[box] (L)  at ($(C2)+(0.9,-0.95)$) {Shared latent $L$};
\node[note] at ($(L)+(0,-0.55)$) {pre-shared, before round};
\node[box] (W) at ($(C3)+(0,-0.95)$) {Shared state, e.g.\ $|W_N\rangle$};

% Agents
\node[sbox] (A11) at ($(C1)+(-1.8,-2.05)$) {Agent 1};
\node[sbox] (A12) at ($(C1)+(0,-2.05)$) {Agent 2};
\node[note] (A13) at ($(C1)+(1.8,-2.05)$) {$\cdots$ Agent $N$};

\node[sbox] (A21) at ($(C2)+(-1.8,-2.05)$) {Agent 1};
\node[sbox] (A22) at ($(C2)+(0,-2.05)$) {Agent 2};
\node[note] (A23) at ($(C2)+(1.8,-2.05)$) {$\cdots$ Agent $N$};

\node[sbox] (A31) at ($(C3)+(-1.8,-2.05)$) {Agent 1};
\node[sbox] (A32) at ($(C3)+(0,-2.05)$) {Agent 2};
\node[note] (A33) at ($(C3)+(1.8,-2.05)$) {$\cdots$ Agent $N$};

\node[note] at ($(C2)+(0,-2.58)$) {no communication, no signaling};

% Intel channels
\node[sbox] (N11) at ($(C1)+(-1.8,-3.25)$) {Intel $p$};
\node[sbox] (N12) at ($(C1)+(0,-3.25)$) {Intel $p$};
\node[note] (N13) at ($(C1)+(1.8,-3.25)$) {$\cdots$};

\node[sbox] (N21) at ($(C2)+(-1.8,-3.25)$) {Intel $p$};
\node[sbox] (N22) at ($(C2)+(0,-3.25)$) {Intel $p$};
\node[note] (N23) at ($(C2)+(1.8,-3.25)$) {$\cdots$};

\node[sbox] (N31) at ($(C3)+(-1.8,-3.25)$) {Intel $p$};
\node[sbox] (N32) at ($(C3)+(0,-3.25)$) {Intel $p$};
\node[note] (N33) at ($(C3)+(1.8,-3.25)$) {$\cdots$};

\node[note] at ($(C2)+(0,-3.82)$) {local, independent perturbations};

% Joint distribution output
\node[box] (P1) at ($(C1)+(0,-4.70)$) {$P(a_1,\ldots,a_N)$};
\node[box] (P2) at ($(C2)+(0,-4.70)$) {$P(a_1,\ldots,a_N)$};
\node[box] (P3) at ($(C3)+(0,-4.70)$) {$P(a_1,\ldots,a_N)$};

\node[note] at ($(C2)+(0,-5.20)$) {no utilities, no equilibria};

% Diagnostics
\node[box] (M) at ($(C2)+(0,-6.25)$) {Diagnostics (payoff-free): APMI,\ TC,\ $\Delta \mathcal{M}$};
\node[note] at ($(M)+(0,-0.65)$) {relative to best classical baseline};

% Arrows
\draw[arr] (L) -- (A21);
\draw[arr] (L) -- (A22);
\draw[arr] (W) -- (A31);
\draw[arr] (W) -- (A32);
\draw[black!55, line width=0.5pt] ($(none)+(-0.85,-0.15)$) -- ($(none)+(0.85,0.15)$);
\draw[black!55, line width=0.5pt] ($(none)+(-0.85,0.15)$) -- ($(none)+(0.85,-0.15)$);
\draw[arr] (A11) -- (N11);
\draw[arr] (A12) -- (N12);
\draw[arr] (A21) -- (N21);
\draw[arr] (A22) -- (N22);
\draw[arr] (A31) -- (N31);
\draw[arr] (A32) -- (N32);
\draw[arr] (N12) -- (P1);
\draw[arr] (N22) -- (P2);
\draw[arr] (N32) -- (P3);
\draw[darr] (H1) -- (N12);
\draw[darr] (H2) -- (N22);
\draw[darr] (H3) -- (N32);
\draw[arr] (P2) -- (M);
\draw[arr,opacity=0.40] (P1) -- (M);
\draw[arr,opacity=0.40] (P3) -- (M);

\end{tikzpicture}
\caption{Schematic of the Hidden-Field Coordination (HFC) model and strategy families. Each round samples a latent hidden field $F$ that remains unobserved by agents and influences only the local intel perturbations (rate $p$). Agents act without communication; correlation arises solely from a pre-shared resource established prior to the round. We compare independent sampling, classical shared-latent coordination (shared variable $L$), and quantum-correlated coordination (shared state (e.g.\ $|W_N\rangle$)). The analyzed object is the empirical joint-action distribution $P(a_1,\ldots,a_N)$, from which payoff-free diagnostics (APMI, total correlation TC, and differentials $\Delta \mathcal{M}$ relative to the best classical baseline) are computed.}
\label{fig:hfc_schematic}
\end{figure*}
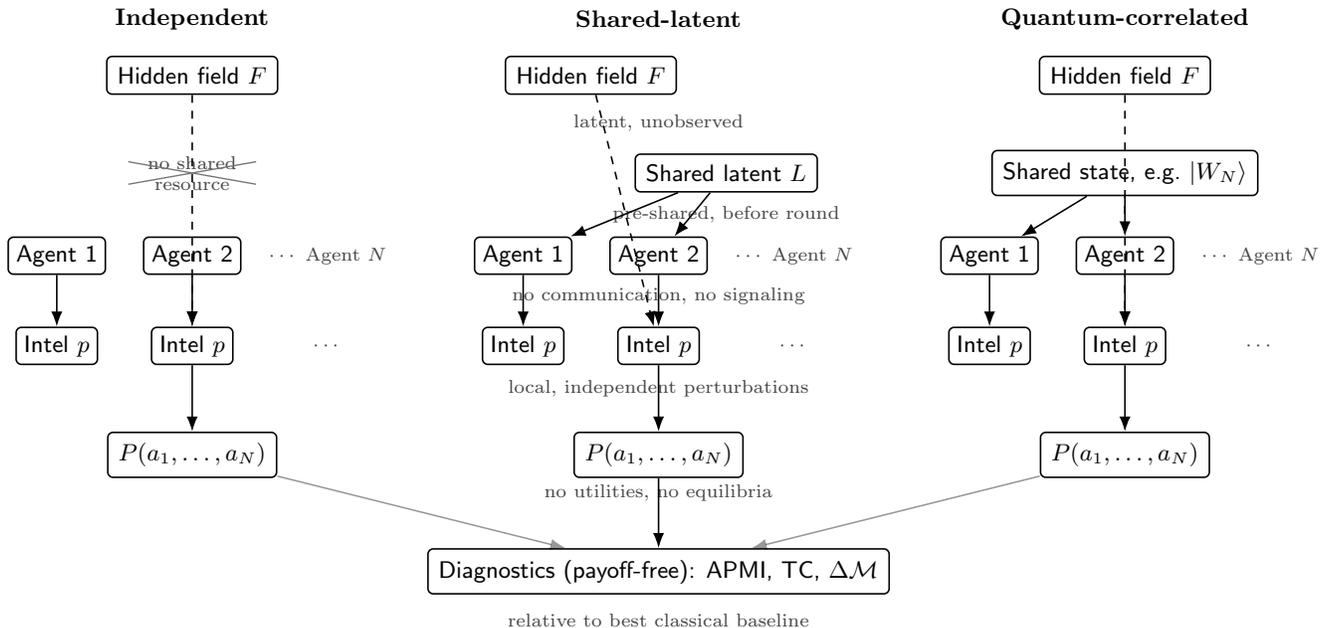

The Hidden-Field Coordination (HFC) model was designed with exactly this separation in mind. Each round opens with an exogenous latent ``field'' that no agent observes and that acts only through the local intel channel, so whatever correlation appears between agents comes from the pre-shared resource and nowhere else. The classical arm supplies a common-cause baseline, the quantum arm an entangled one. The map from measurement outcome to action is fixed in advance and identical for every agent, which rules out the possibility that a difference in the joint statistics is really a difference in how cleverly each strategy was tuned to the task. Figure~\ref{fig:hfc_schematic} schematically summarizes the HFC generative process.

We characterize coordination using the \emph{average pairwise mutual information} (APMI) and the \emph{total correlation} (TC) \cite{Shannon1948, CoverThomas2006, Watanabe1960, walczak2009total}. To isolate the geometric separation between strategies, we report \emph{differentials} relative to the strongest classical baseline. As shown in our results, a negative differential in these metrics indicates that the quantum strategy has sacrificed raw correlation magnitude, the ``alignment trap,'' to achieve a collision-suppressing geometry. This perspective aligns with recent views of coordination and shared randomness as resources in multi-agent quantum settings \cite{Natur2025}.

We implement the entanglement-mediated strategy using $W$ states, which are natural candidates for distributed coordination because they retain entanglement under single-particle loss \cite{Ozdemir2011, Bugu2013A, Ozaydin2014A,HUANG2024129492, bugu2020preparing, Thapa2025, Horodecki2009Review, Guhne2009Multipartite, MA2025, Dur2000, Zhu2024, Huber2010Robustness, park2025_W_entangled}. Our focus is therefore a minimal model of multipartite coordination under realistic NISQ-era constraints \cite{NielsenChuang2010, preskill2018quantum, Brandao2017QuantumDeFinetti, Harrow2017Sampling}.

In this work, we identify a constraint on classical coordination under symmetry. For any anonymous strategy with shared classical randomness and no communication, we prove that the pairwise coincidence cannot fall below the inverse alphabet size, however the shared randomness is used. We show that quantum-correlated strategies based on multipartite entanglement can violate this constraint and produce joint-action distributions that combine nontrivial dependence with suppressed collisions. This separation appears without any advantage in expected payoff, indicating that entanglement reshapes the geometry of coordination beyond what is visible at the level of expectation values. Section~\ref{sec:model} states and proves the bound, Sec.~\ref{sec:results} locates the quantum operating point relative to it, and Sec.~\ref{subsec:scaling} delimits the $N < 2M$ window in which the violation survives.

% ==============================================================================
% FIGURE 1: SCHEMATIC
% ==============================================================================
% ==============================================================================
% SECTION II: MODEL
% ==============================================================================
\section{Hidden-Field Coordination Model}
\label{sec:model}

We introduce the Hidden-Field Coordination (HFC) model as a minimal probabilistic framework for analyzing correlation structure in decentralized multi-agent systems under partial information.
The defining feature of the model is an exogenous, round-by-round random latent variable that reshapes which joint action patterns are penalized or favored, without being directly observable by the agents.
The model is deliberately constructed to separate correlation structure from payoff optimization, communication, or equilibrium reasoning.

\subsection{Generative structure}

Consider $N$ agents, each selecting an action from a finite alphabet
$\mathcal{A}=\{1,\dots,M\}$.
Each round proceeds according to the following generative process.

At the start of the round, a hidden field $F$ is sampled from a fixed distribution.
In the instantiations studied here, $F$ corresponds to a random selection of $K$ unfavorable targets among the $M$ available actions.
The field is exogenous, fixed throughout the round, and inaccessible to all agents.

Each agent $i$ produces an initial action proposal $\tilde a_i \in \mathcal{A}$ according to a strategy-dependent sampling rule.
Agents do not communicate, do not observe each other’s actions, and do not observe $F$.
Any correlation between agents must therefore arise from shared resources established prior to the round.

A local ``intel'' process then perturbs each agent’s proposed action independently with probability $p \in [0,1]$. 
The intel process is memoryless and depends only on the agent’s current proposal and its noisy local interaction with the hidden field, with no dependence on past rounds or other agents.
After this perturbation, each agent outputs a final action $a_i$. No utility function or equilibrium concept is invoked; the analysis focuses purely on the resulting empirical joint-action distribution $P(a_1,\dots,a_N)$.

\subsection{Information-theoretic observables}

To characterize coordination independently of outcomes, we analyze $P(\mathbf{a})$ using:

\paragraph{Average pairwise mutual information (APMI):}
\begin{equation}
\mathrm{APMI}
= \frac{2}{N(N-1)} \sum_{i<j} I(a_i;a_j),
\end{equation}
summarizing pairwise structure across the team \cite{CoverThomas2006}.

\paragraph{Total correlation (TC):}
\begin{equation}
\mathrm{TC}(\mathbf{a})
= \sum_{i=1}^{N} H(a_i) - H(\mathbf{a}),
\end{equation}
which captures genuinely multipartite dependence \cite{Watanabe1960}. TC vanishes if and only if the joint distribution factorizes.

\subsection{Classical coordination constraint}

We restrict attention to anonymous, index-equivariant strategies with shared classical randomness and no communication. Here ``anonymous'' means agents have no pre-assigned identities, and ``index-equivariant'' means all agents apply the same local response rule. We regard this as the natural classical baseline under the symmetry constraints of the task: any correlation must arise from a shared latent resource established before the round, not from communication or externally injected role labels.
We compare three strategy family: \emph{independent classical sampling}, where each agent acts without any shared resource; \emph{shared-latent classical coordination}, where agents share a common classical variable $L$ and each follows it with probability $q$; and \emph{quantum-correlated 
coordination}, where agents share an $N$-partite $W$ state and apply the 
\emph{Spontaneous Leader Election} protocol locally. Full implementation 
details are given in Sec.~\ref{sec:strategies}.
\medskip
\noindent\textbf{Proposition 1 (Classical coincidence lower bound).} Consider any classical strategy in which anonymous agents share an arbitrary classical random variable $\Lambda$ (of any alphabet, with any distribution), each agent applies the same stochastic response rule to $\Lambda$ using private, independent local randomness, and no communication takes place. Then the pairwise coincidence satisfies
\begin{equation}
P(a_i = a_j) \ge \frac{1}{M},
\label{eq:prop1}
\end{equation}
with equality if and only if the conditional single-agent action distribution is uniform for almost every value of $\Lambda$.

\noindent\emph{Proof.} Anonymity and the absence of communication imply that, conditioned on $\Lambda = \lambda$, the agents' actions are independent and identically distributed with some conditional distribution $p_\lambda$ on $\{1,\dots,M\}$. Hence
\begin{equation}
P(a_i = a_j) = \mathbb{E}_\Lambda\!\left[\sum_{a=1}^{M} p_\Lambda(a)^2\right] \ge \mathbb{E}_\Lambda\!\left[\frac{1}{M}\right] = \frac{1}{M},
\end{equation}
where the inequality is Cauchy--Schwarz, saturated only by the uniform distribution. The hidden field is covered by the same argument: since the intel channel acts independently on each agent conditioned on the field realization $f$, the pair $\Lambda' = (\Lambda, F)$ is itself a classical common cause, and the bound applies to the executed actions $a_i$. The one-parameter shared-latent family of Sec.~\ref{sec:strategies} is a special case; within that family the coincidence takes the closed form $\mathrm{Coin}(q) = q^2(1-1/M) + 1/M$, minimized at $q=0$. \hfill$\square$

\subsection{Differential classical baselines}

We also quantify collisions via the \emph{average pairwise coincidence}
\begin{equation}
\mathrm{Coin}
= \frac{2}{N(N-1)} \sum_{i<j} \Pr[a_i=a_j].
\end{equation}
For completeness, one may also define the \emph{global} collision probability $\Pr[a_1=\cdots=a_N]$; throughout this paper, the term ``collision'' refers to the pairwise coincidence functional $\mathrm{Coin}$ unless explicitly stated otherwise.

We report observables relative to the strongest classical baseline within the shared-latent family:
\begin{equation}
\Delta \mathcal{M}
= \mathcal{M}_{\mathrm{quant}}
- \max\Bigl\{\mathcal{M}_{\mathrm{ind}},\ \max_{q\in[0,1]} \mathcal{M}_{\mathrm{shared}}(q)\Bigr\}.
\end{equation}
Within this restricted symmetric baseline, increasing $\mathcal{M}$ is typically achieved by increasing action-alignment (copying), so \emph{negative} differentials quantify separation between the alignment-based classical optimum (over $q$) and the collision-suppressing quantum structure.
This allows us to study quantum correlations as \emph{distributional objects}, disentangled from payoff optimization or signaling.
% ==============================================================================
% SECTION III: RESULTS
% ==============================================================================
\section{Results}
\label{sec:results}

We report how entanglement-mediated sampling in the Hidden-Field Coordination (HFC) model modifies the \emph{distributional structure} of multi-agent coordination.
Throughout, we compare three strategy families: (i) independent classical sampling, (ii) a shared-latent classical coordination model (optimized over $q$ for differential metrics), and (iii) a quantum-correlated strategy implemented by mapping measurement outcomes of an $N$-qubit $W$ state to local actions via the \emph{Spontaneous Leader Election} protocol.
All strategies are evaluated under identical hidden-field realizations and identical local intel perturbations; the quantum strategy additionally includes gate-level depolarizing noise of strength $\lambda$ in the $W$-state preparation circuit.
No payoff is computed anywhere below. Every quantity reported is a functional of the joint action distribution alone.

\subsection{Joint-action signatures}
\label{subsec:jointmaps}

\begin{figure*}[t] 
  \centering
  % Top row (N=3)
  \includegraphics[width=\textwidth]{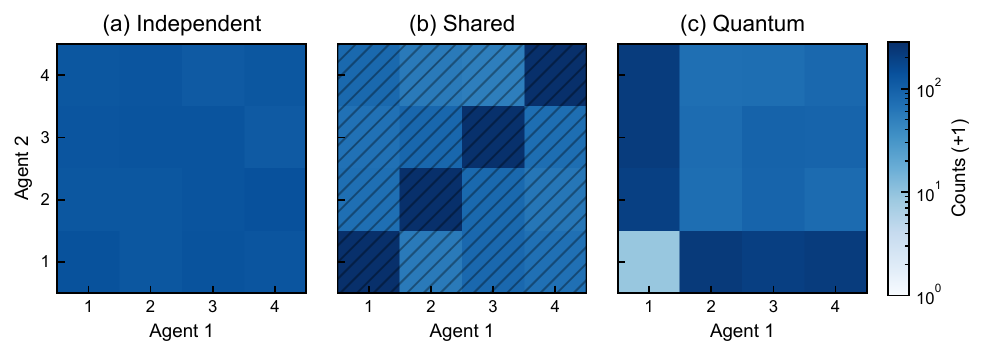}
  
  % Add a little vertical space between rows
  \vspace{0.2cm} 
  
  % Bottom row (N=5)
  \includegraphics[width=\textwidth]{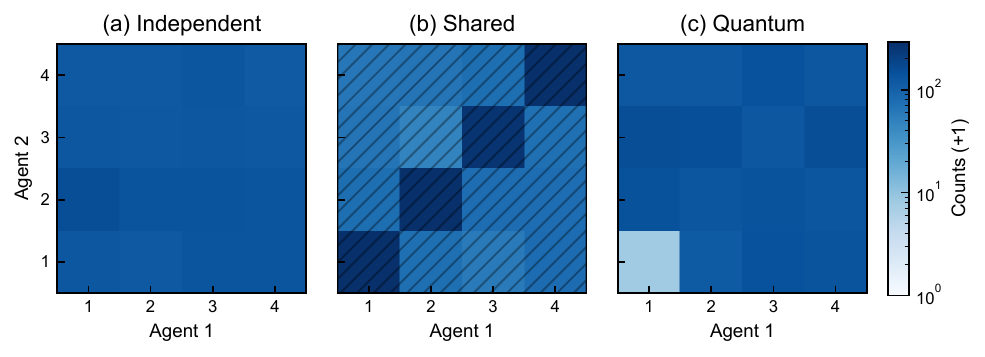}
  
  \caption{
    Representative joint-action maps for team sizes $N=3$ (top) and $N=5$ (bottom).
    The shared-latent model (middle panels, diagonal strips, $q=0.7$) shows concentration
    along the $a_1 = a_2$ diagonal, indicating coordination achieved through action alignment (copying).
    In contrast, the quantum strategy (right panels) displays an ``L-shaped'' relational structure. 
    Probabilities are concentrated where one agent is the Leader ($a=1$) and others are Followers ($a \neq 1$), 
    effectively distributing probability mass away from the collision point at $(1,1)$.
  }
  \label{fig:jointmaps}
\end{figure*}

Figure~\ref{fig:jointmaps} shows empirical two-agent joint-action distributions.
The shared-latent classical strategy (middle panels) concentrates probability along the diagonal ($a_i = a_j$), reflecting correlations dominated by action equality (copying). Within the restricted shared-latent baseline family defined in Sec.~\ref{sec:model}, increasing dependence proceeds primarily through this alignment mechanism.

In contrast, the quantum strategy (right panels) exhibits a distinct collision-suppressing geometry. High probability mass is concentrated in off-diagonal configurations where exactly one agent assumes the Leader role ($a=1$).
This structure demonstrates how the $W$ state's spontaneous symmetry breaking allows agents to coordinate on unequal roles without pre-assigned indices. The quantum map shows a ``hole'' at the $(1,1)$ coordinate compared to the classical case, demonstrating that entanglement-mediated sampling can coordinate roles while suppressing collisions on the Leader action, below the level attainable by any anonymous classical common-cause strategy.
\FloatBarrier

\subsection{Coordination geometry in the \texorpdfstring{$(\mathrm{APMI}, \mathrm{Coincidence})$}{(APMI, Coincidence)} plane}
\label{subsec:geometry}

\begin{figure}[t]
  \centering
  \includegraphics[width=\columnwidth]{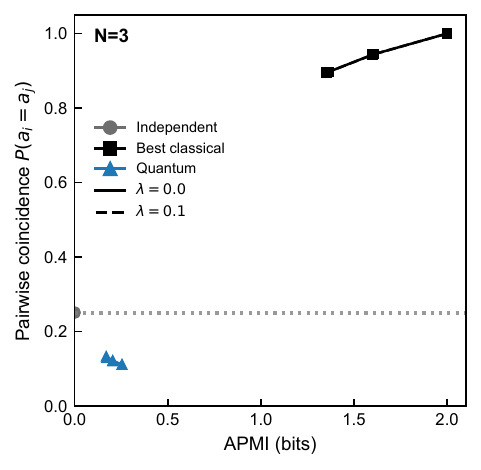}

  \par\medskip

  \includegraphics[width=\columnwidth]{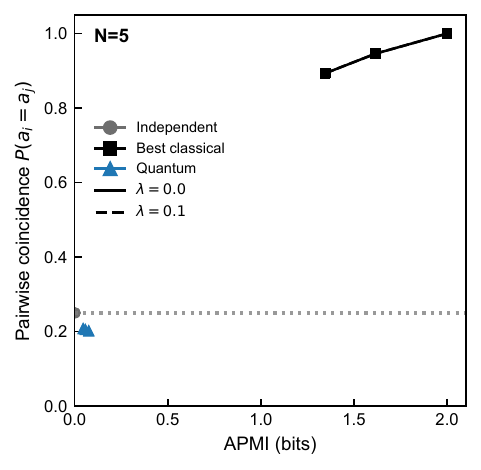}
 \caption{
  Coordination geometry trails for $N=3$ (top) and $N=5$ (bottom).
  The shared-latent classical model (black squares) traces a path toward high coincidence, confirming that the restricted shared-latent classical baseline increases correlation (APMI) primarily through alignment.
  The quantum strategy (blue triangles) occupies a distinct region characterized by stable APMI but low coincidence, sitting below the grey dotted line at $1/M$, which is simultaneously the independent baseline and the classical coincidence lower bound of Proposition~1. The quantum strategy therefore lies in a region unreachable by any anonymous classical common-cause strategy.
  }
  \label{fig:geometry}
\end{figure}

%% feasibility figure
\begin{figure}[t]
  \centering

  \includegraphics[width=\columnwidth]{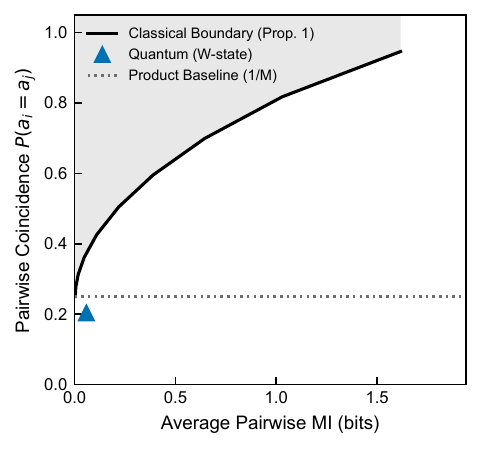}
\caption{Feasible region of symmetric classical coordination in the (APMI, Coincidence) plane. The curve is obtained by sweeping the shared-latent parameter $q \in [0,1]$ and traces the dependence--collision trade-off realized by that family; the horizontal level $1/M$ marks the coincidence lower bound of Proposition~1, which applies to every anonymous classical common-cause strategy. Classical strategies outside the shared-latent family may realize other combinations of APMI and coincidence, but none can fall below $1/M$. The quantum operating point ($\mathrm{Coin}=0.205\pm0.001$, $\mathrm{APMI}=0.057\pm0.001$, mean $\pm$ s.e.m.\ across 8 replicates) lies well below the classical lower bound $1/M=0.25$, with the observed separation far exceeding the estimated statistical uncertainty. This confirms a statistically significant violation of the coincidence lower bound established in Proposition~1.}
\label{fig:feasible_region}
\end{figure}

We project each strategy into a two-dimensional plane spanned by average pairwise coincidence ($\mathrm{Coin}$) and average pairwise mutual information ($\mathrm{APMI}$).
Figure~\ref{fig:geometry} shows coordination ``trails'' as the intel rate $p$ is varied.
The shared-latent classical strategy moves toward the top-right corner, demonstrating that increasing coordination (APMI) in classical systems raises the probability of action identity within this family.

The quantum strategy behaves quite differently. It occupies a distinct region in the bottom-left and often dips \emph{below} the coincidence level of independent agents.
Within the classical baseline family, reducing coincidence requires weakening the shared-latent component, which drives APMI toward zero.
The entanglement-mediated mapping, however, maintains nontrivial dependence while occupying a lower-coincidence region.
The separation is not uniform in $N$: the coincidence gap closes once $N \ge 2M$, as Eq.~(\ref{eq:coinW}) below makes explicit. Within the window it survives, the departure from alignment-based dependence is a property of the $W$ resource rather than of any particular noise setting. This separation is made explicit in Fig.~\ref{fig:feasible_region}, which shows that the quantum strategy occupies a region below the classical coincidence lower bound.
\FloatBarrier

\subsection{Global dependence across intel rate and depolarizing noise}
\label{subsec:globalTC}

Pairwise metrics often obscure genuinely multipartite structure. To capture global dependence, we compute the total correlation (TC). We compare strategies using the differential $\Delta \mathrm{TC}(p,\lambda)$ relative to the best-performing classical baseline.

\begin{figure*}[tb]
  \centering
  \includegraphics[width=0.49\textwidth]{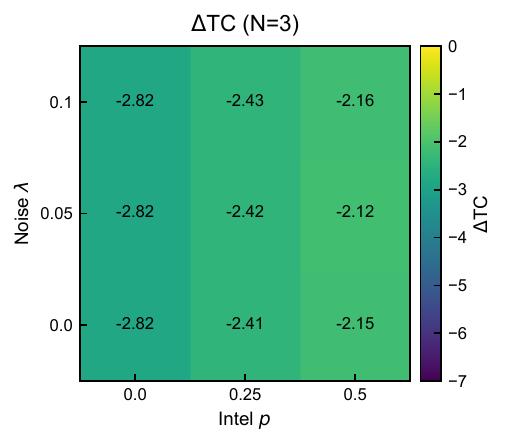}
  \includegraphics[width=0.49\textwidth]{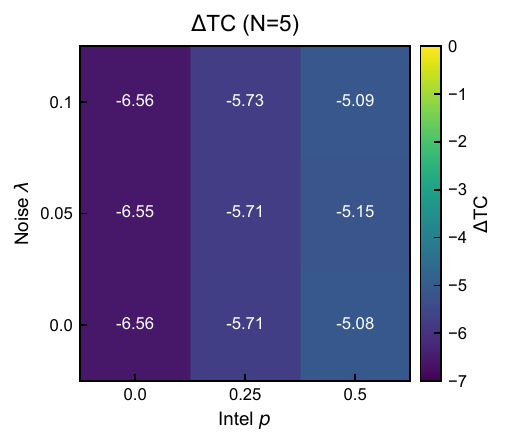}
  \caption{
  Differential Total Correlation $\Delta \mathrm{TC}$ for $N=3$ and $N=5$.
  Consistently negative values across the noise ($\lambda$) and intel ($p$) sweep confirm that the classical baseline achieves higher raw dependence via alignment. 
  The quantum strategy uses the $W$ state to generate relational correlations, sacrificing total shared information to remain in a low-collision regime.
  }
  \label{fig:deltaTC}
\end{figure*}

Figure~\ref{fig:deltaTC} reports $\Delta \mathrm{TC}$ for $N=3$ and $N=5$.
Across the explored parameter region, $\Delta \mathrm{TC}$ remains negative.
The reason is visible in the $q$-sweep of Appendix~\ref{app:qsweep}: the shared-latent model reaches its maximum TC only as $q \to 1$, where $a_1 = \dots = a_N$. While this yields a higher raw magnitude of shared bits, it results in maximal local collisions.
The quantum strategy never approaches that limit, so it forfeits those bits. What $\Delta\mathrm{TC} < 0$ measures is the size of that forfeit, not a deficit in coordination: the shared bits the classical baseline gains are precisely the ones that put both agents on the same action.

\begin{figure*}[tb]
  \centering
  \includegraphics[width=0.49\textwidth]{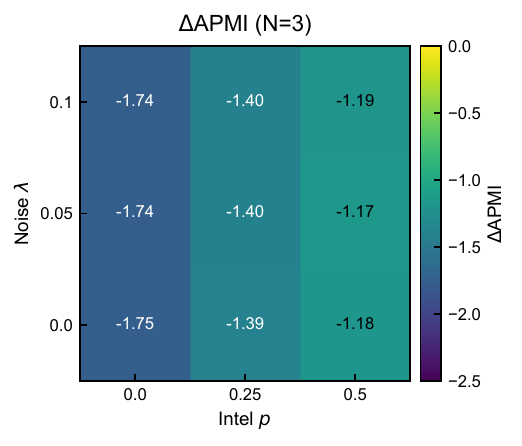}
  \includegraphics[width=0.49\textwidth]{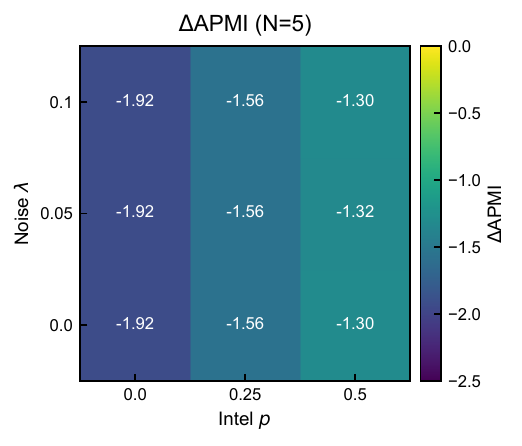}
\caption{Pairwise-structure differential $\Delta \mathrm{APMI}$ between the quantum strategy and the best classical baseline across intel rate $p$ and depolarizing strength $\lambda$, shown for $N=3$ (left) and $N=5$ (right). Here $\Delta \mathrm{APMI} = \mathrm{APMI}_{\mathrm{quant}} - \max\!\left(\mathrm{APMI}_{\mathrm{ind}}, \max_{q\in[0,1]}\mathrm{APMI}_{\mathrm{shared}}(q)\right)$ evaluated \emph{locally at each} $(p,\lambda)$. The consistently negative differentials highlight that the quantum strategy does not aim to maximize raw information-theoretic magnitude. The near-identical appearance of the two panels reflects the comparatively stable $\Delta\mathrm{APMI}$ across team sizes ($N=3$: $-1.75$ to $-1.17$\,bits; $N=5$: $-1.92$ to $-1.30$\,bits), consistent with the scaling result in Fig.~\ref{fig:scaling}. The primary advantage of the quantum strategy lies in the \emph{geometric distribution} of those correlations (lower coincidence) relative to the classical baseline, which is forced into action-alignment to achieve comparable APMI levels.}
  \label{fig:deltaAPMI}
\end{figure*}

Similarly, Fig.~\ref{fig:deltaAPMI} shows the differential in APMI, which also remains negative.
The same holds pairwise: the value of the quantum resource lies in its \emph{geometry} (the avoidance of collisions seen in Figure~\ref{fig:geometry}) rather than its absolute information-theoretic magnitude.
\FloatBarrier

\subsection{Scaling of geometric separation}
\label{subsec:scaling}

\begin{figure}[t]
  \centering
  \includegraphics[width=\columnwidth]{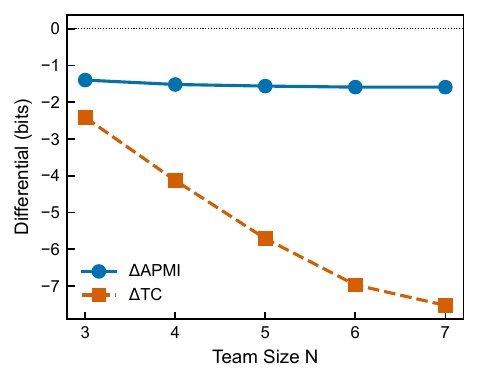}
\caption{
Scaling of correlation differentials at $(p=0.25, \lambda=0)$ for $N\in\{3,4,5,6,7\}$.
Over the tested range, $\Delta\mathrm{TC}(N)$ becomes increasingly negative while $\Delta\mathrm{APMI}(N)$ remains comparatively stable, indicating that the separation between the quantum strategy and the best-performing classical baseline within the restricted shared-latent family is more pronounced in a global dependence metric than in a pairwise one. Circles (blue, solid) denote $\Delta\mathrm{APMI}$; squares (vermillion, 
dashed) denote $\Delta\mathrm{TC}$.
}
  \label{fig:scaling}
\end{figure}

We examine how these signatures vary with team size at a fixed operating point ($p=0.25$, $\lambda=0$) by extending the analysis to $N\in\{3,4,5,6,7\}$.
Figure~\ref{fig:scaling} plots the differentials $\Delta \mathrm{APMI}(N)$ and $\Delta\mathrm{TC}(N)$.

Over the tested range, we observe a systematic trend: $\Delta\mathrm{TC}(N)$ becomes increasingly negative as $N$ grows, while $\Delta\mathrm{APMI}(N)$ remains comparatively stable.
This indicates that the gap in raw information-theoretic magnitude between the quantum strategy and the best-performing classical baseline within the restricted shared-latent family grows more strongly in a global metric (TC) than in a pairwise one (APMI).
Together with the coincidence diagnostics, this behavior is consistent with a structural distinction that is primarily multipartite: the quantum strategy maintains a non-alignment-based relational profile under the fixed local mapping, whereas the restricted shared-latent baseline can increase dependence primarily by shifting probability mass toward action alignment.

The coincidence violation itself has a finite window in team size. For the ideal noiseless protocol, exactly one agent occupies the Leader role, and the pairwise coincidence evaluates to
\begin{equation}
\mathrm{Coin}_W = \frac{1 - 2/N}{M-1},
\label{eq:coinW}
\end{equation}
which lies below the classical bound $1/M$ of Proposition~1 if and only if $N < 2M$. For the alphabet size $M=4$ used here, the violation therefore persists up to $N = 7$ and closes at $N = 8$. Equation~(\ref{eq:coinW}) describes the noiseless protocol with no intel channel, giving $1/9 \approx 0.111$ at $N=3$ and $0.200$ at $N=5$. The simulated values at $p=0.25$ are $0.122 \pm 0.002$ and $0.205 \pm 0.001$; the excess over the noiseless values reflects the intel channel mixing executed actions toward uniform, and in neither case does it close the gap to the bound $0.25$. The window widens linearly with the action alphabet, so the coincidence violation is a small-team, large-alphabet effect; by contrast, the global separation $\Delta\mathrm{TC}(N)$ in Fig.~\ref{fig:scaling} grows monotonically over the tested range and is not tied to this window.

\FloatBarrier
% ==============================================================================
% SECTION IV: METHODS
% ==============================================================================
\section{Methods}
\label{sec:methods}

\subsection{Hidden-Field Coordination model}
\label{subsec:model_methods}

We study a Hidden-Field Coordination (HFC) task involving $N$ agents and $M$ discrete actions.
At the beginning of each round, an external hidden field selects a subset $F \subset \{1,\dots,M\}$ of cardinality $K$.
This assignment is random, exogenous, and unobserved by the agents.
The hidden field do not constitute a player and does not respond to agent actions.

Each agent $i$ produces an initial action proposal $\tilde a_i \in \{1,\dots,M\}$ according to a strategy-dependent sampling rule.
The final executed action $a_i$ may differ from $\tilde a_i$ due to a local intel perturbation process defined below.

Agents do not communicate and have no access to the hidden-field realization.
All strategies are evaluated under identical hidden-field sequences and intel realizations to ensure controlled comparisons; the quantum strategy additionally includes gate-level depolarizing noise of strength $\lambda$ in the state-preparation circuit.
\FloatBarrier

\subsection{Intel perturbation channel}
\label{subsec:intel_methods}

Partial interaction with the hidden field is modeled as a local, agent-wise stochastic channel applied after each strategy produces an initial proposal $\tilde a_i\in\{1,\dots,M\}$.
Each round samples a hidden field as an indicator vector
$f \in \{0,1\}^M$ with exactly $K$ defended actions, i.e., $\sum_{t=1}^M f_t = K$.
The field is unobserved by agents and fixed throughout the round.

The intel channel acts independently on each agent $i$ and depends only on $(\tilde a_i,f)$.
With probability $p$ (the intel rate), agent $i$ performs an ``inspection'': it draws an index $T\sim\mathrm{Unif}\{1,\dots,M\}$ and observes a noisy defended-bit
\begin{equation}
\hat f_T = f_T \oplus B,
\qquad
B\sim\mathrm{Bernoulli}(\epsilon),
\label{eq:intel_noisy_bit_methods}
\end{equation}
where $\oplus$ denotes XOR and $\epsilon$ is the intel-flip probability.
If the inspected action coincides with the agent's proposal and the noisy bit indicates ``defended,''
\begin{equation}
\tilde a_i = t \ \ \text{and}\ \ \hat f_t = 1,
\end{equation}
then with probability $v$ (the veto probability) the agent redirects its action uniformly to one of the remaining $M-1$ actions.
Thus $v$ controls the conditional probability of redirection given that inspection is performed, the inspected index matches the proposal, and the noisy defended-bit reports ``defended.'' 
\begin{equation}
a_i \sim \mathrm{Unif}\big(\{1,\dots,M\}\setminus\{t\}\big).
\label{eq:intel_redirect_methods}
\end{equation}
In all other cases (no inspection, or inspection that does not trigger a veto), the executed action remains unchanged, $a_i=\tilde a_i$.

The intel channel is memoryless across rounds and conditionally independent across agents given the hidden field, and it introduces no signaling or shared randomness between agents.

In all simulations presented in this work, we use $M = 4$ actions and $K = 2$ defended actions. The intel parameter $p$ is varied over $\{0, 0.25, 0.5\}$ and the depolarizing noise strength $\lambda$ over $\{0, 0.05, 0.1\}$, as indicated in the corresponding figures. The veto probability is fixed at $v = 0.9$.
\FloatBarrier

\subsection{Strategy Families}
\label{sec:strategies}

We compare three families of decentralized strategies that differ in their underlying correlation resources.

\paragraph{Independent classical sampling.}
Each agent samples $\tilde a_i$ independently and uniformly from $\mathcal{A}$. This provides a baseline with no shared correlation resource where the joint distribution factorizes.

\paragraph{Shared-latent classical coordination.}
Agents share a classical latent variable $L \in \mathcal{A}$. Conditioned on $L$, each agent outputs $L$ with probability $q$ and samples uniformly from $\mathcal{A}$ with probability $1-q$. This one-parameter family is the restricted symmetric classical baseline used throughout the paper (the shared latent takes values in the action alphabet and all agents use the same conditional response). As $q \to 1$, this strategy converges to the ``copy'' limit ($a_1 = \dots = a_N$), maximizing total correlation within the ideal shared-latent family while also maximizing coincidence. For the numerical differential metrics, the shared-latent baseline was evaluated on a uniform grid containing $q=1$, and the largest sampled value was used at each operating point; in all evaluated configurations, including a finer 41-point scan of the post-channel observables, the maximum occurred at $q=1$.

\paragraph{Quantum-correlated coordination.}
Agents share an $N$-partite entangled $W$ state prior to the round:
\begin{equation}
\lvert W_N \rangle = \frac{1}{\sqrt{N}} \sum_{k=1}^{N} \lvert 0\cdots 1_k \cdots 0 \rangle,
\end{equation}
chosen for its robustness to local noise and particle loss \cite{Zhu2024, chaves2010robustness}. To ensure decentralized coordination without signaling, agents use a \emph{Spontaneous Leader Election} protocol. Each agent $i$ performs a local projective measurement in the computational basis ($b_i \in \{0, 1\}$). If $b_i = 1$ (excitation), the agent adopts a ``Leader'' role and selects action $1$. If $b_i = 0$ (ground), the agent adopts a ``Follower'' role and samples an action uniformly at random from the set $\{2, \dots, M\}$. This mapping is strictly local and index-equivariant; no global information regarding the total number of excitations is used by the agents.

\subsection{Noise and Protocol Robustness}
\label{subsec:noise_methods}

In the quantum strategy, we model decoherence by attaching depolarizing errors of strength $\lambda$ to every gate of the transpiled state-preparation circuit (single-qubit errors on the basis gates, two-qubit errors on entangling gates), so the effective noise grows with the depth of the $W$-state preparation. The \emph{Spontaneous Leader Election} protocol defined in 
Section~\ref{sec:strategies} enforces the no-communication constraint under these NISQ-era conditions \cite{preskill2018quantum}.

Under ideal conditions, the $W$ state ensures exactly one agent measures $b_i = 1$, resulting in a perfectly coordinated occupation of the Leader role. However, if decoherence or gate errors result in multiple excitations ($\sum b_i > 1$) or none at all ($\sum b_i = 0$), the agents cannot ``re-negotiate'' or communicate to resolve the discrepancy. They may unintentionally collide on the Leader action or leave it unoccupied. This strict adherence to local measurement outcomes ensures that no hidden communication channels are present, providing a stringent benchmark against classical decentralized baselines.
% \subsection{Simulation procedure}
% \label{subsec:simulation_methods}

% Simulations are performed using the \texttt{Qiskit Aer} \cite{cross2018ibm} backend.
% For each configuration $(N,p,\lambda)$, we generate $R=2000$ independent rounds per replicate and average over $8$ statistically independent replicates.
% Hidden-field realizations, intel perturbations, and quantum transpilation seeds are derived deterministically from hashed configuration identifiers.
% All reported observables are computed from the aggregated joint action arrays $\mathbf{a} = (a_1,\dots,a_N)$.

% \subsection{Reproducibility}
% \label{subsec:reproducibility_methods}

% All figures are generated using a deterministic pipeline. Fixed seeds control every stochastic component, including hidden-field realizations, intel perturbations, circuit transpilation, and quantum shot sampling, and the scripts reproduce the reported values and figures under the documented software environment. In addition, the classical bound of Proposition~1 was stress-tested numerically by adversarial minimization of the pairwise coincidence over anonymous shared-randomness strategies of varying latent-alphabet size, both with and without the intel channel; no classical strategy fell below $1/M$.
% ==============================================================================
% SECTION V: DISCUSSION
% ==============================================================================
\section{Discussion}
\label{sec:discussion}

The Hidden-Field Coordination (HFC) model introduced in this work was designed to isolate correlation structure in decentralized multi-agent systems independently of payoff optimization, equilibrium selection, or communication.
Within this controlled setting, our results demonstrates a qualitative geometric distinction between classical alignment-based strategies and entanglement-mediated coordination.

The two strategy families end up in different parts of the (APMI, Coin) plane, and they get there by different routes.
As shown in the geometric trails (Fig.~\ref{fig:geometry}), the shared-latent classical baseline maximizes correlation only by driving the system toward high action coincidence (the diagonal).
The result is a fundamental classical trade-off: agents can be highly correlated (by copying) or collision-suppressing (by randomizing), but not both.
The $W$ strategy sits outside that trade-off entirely. Proposition~1 fixes $1/M$ as a floor on coincidence for every anonymous classical common-cause strategy, and Fig.~\ref{fig:feasible_region} places the simulated quantum point at $0.205 \pm 0.001$ against a floor of $0.25$ --- roughly fifty standard errors below it. That margin is not permanent: Eq.~(\ref{eq:coinW}) puts the crossing at $N = 2M$, so at $M=4$ the violation is a property of teams of seven or fewer. The protocol preserves rule symmetry (all agents follow the same conditional logic) while allowing the state itself to spontaneously break symmetry round-by-round. This enables significant global dependence (as evidenced by nonzero Total Correlation in the noiseless limit) while reducing local collisions relative to the classical alignment baseline.

The negative differential total correlation ($\Delta \mathrm{TC}$) observed in Fig.~\ref{fig:deltaTC} quantifies this separation.
It confirms that classical common-cause models achieve higher raw dependence, but only by paying the cost of maximal collision probability.
The quantum strategy sacrifices raw bit-count relative to this ``copying maximum'' to access a collision-suppressing geometry that Proposition~1 shows is inaccessible to any anonymous classical common-cause strategy.

The choice of $W$ states as the primary resource is motivated by their unique structure for decentralized tasks. Unlike GHZ-type states, which concentrate correlations into global phases that are highly fragile to local perturbations, the $W$ state provides a ``relational index'' (the excitation position $k$) that is naturally suited for distributing distinct, coordinated roles among agents.
This allows the quantum strategy to generate nonzero total correlation while maintaining robustness to the local ``intel'' perturbations that would otherwise collapse more rigid entanglement structures.

Entanglement does not give the team more correlation here --- by both metrics it gives less. What it changes is which pairs (APMI, Coin) are reachable at all.
The primary advantage of the quantum resource is its ability to access a collision-suppressing coordination regime that is not achieved by the shared-latent classical baseline family considered here.
Viewed this way, the HFC framework provides a straightforward way to separate exogenous latent structure from coordination mechanisms while comparing correlation differentials against the strongest classical baselines allowed by the model.
This approach complements recent efforts \cite{kao2024scalable, poggi2024measurement} to characterize quantum correlations in networked and multi-agent settings beyond outcome-centric or equilibrium-based analyses.

\section*{Data Availability}
The numerical data underlying all figures in this work, together with the code used to generate them, are available from the corresponding author upon reasonable request.

\bibliography{references}
\end{document}